\documentclass[12pt]{article}

\newcommand{\ltap}{\
  \raise.3ex\hbox{$<$\kern-.75em\lower1ex\hbox{$\sim$}}\ }
\newcommand{\gtap}{\
  \raise.3ex\hbox{$>$\kern-.75em\lower1ex\hbox{$\sim$}}\ }

\makeatletter
\def\myabstract#1{\gdef\@myabstract{#1}}
\def\@myabstract{\@latex@warning@no@line{No \noexpand\myabstract given}}
\def\preprint#1{\gdef\@preprint{#1}}
\def\@preprint{\@latex@warning@no@line{No \noexpand\preprint given}}

\renewcommand\maketitle{\par
  \begingroup
    \renewcommand\thefootnote{\@fnsymbol\c@footnote}%
    \def\@makefnmark{\rlap{\@textsuperscript{\normalfont\@thefnmark}}}%
    \long\def\@makefntext##1{\parindent 1em\noindent
            \hb@xt@1.8em{%
                \hss\@textsuperscript{\normalfont\@thefnmark}}##1}%
    \if@twocolumn
      \ifnum \col@number=\@ne
        \@maketitle
      \else
        \twocolumn[\@maketitle]%
      \fi
    \else
      \newpage
      \global\@topnum\z@   
      \@maketitle
    \fi
 \endgroup
  \setcounter{footnote}{0}%
  \global\let\thanks\relax
  \global\let\maketitle\relax
  \global\let\@maketitle\relax
  \global\let\@thanks\@empty
  \global\let\@author\@empty
  \global\let\@date\@empty
  \global\let\@title\@empty
  \global\let\title\relax
  \global\let\author\relax
  \global\let\date\relax
  \global\let\and\relax
}

\renewcommand\@maketitle{%
  \begin{titlepage}
  \null
\begin{flushright}
\@preprint
\end{flushright}
  \vskip 60\p@%
  \begin{center}%
  \let \footnote \thanks
    {\LARGE \@title \par}%
    \vskip 3em%
    {\large
      \lineskip .75em%
      \begin{tabular}[t]{c}%
        \@author
      \end{tabular}\par}%
    \vskip 1.5em%

  \end{center}%
  \par
  \begin{abstract}
    \@myabstract
    \end{abstract}
    \par\vfill
    \begin{flushleft}
      {\large \@date}
    \end{flushleft}
    \@thanks
\vfil\null
\end{titlepage}}
\makeatother

\begin{document}

\preprint{CERN-TH/97-208\\
BUHEP-97-26}

\title{SCARS ON THE CBR?}
\author{A.G. Cohen$^a$ and A. De
  R\'ujula$^{b,a}$\ \thanks{\tt cohen@bu.edu,
    derujula@nxth21.cern.ch}\\ \\
  \small \sl $^a$Department of Physics, Boston University,
  Boston, MA 02215, USA \\
  \small \sl $^b$Theory Division, CERN,
  1211 Geneva 23, Switzerland \\
  }

\myabstract{We ask whether the universe can be a patchwork consisting of
  distinct regions of matter and antimatter.  In previous work we
  demonstrated that post-recombination matter--antimatter contact near
  regional boundaries leads to an observable (but unobserved)
  gamma-ray flux for domain sizes of less than a few thousand Mpc,
  thereby excluding such domains. In this paper we consider the {\it
    pre}-recombination signal from domains of larger size.}

\date{September, 1997}
\maketitle

We recently studied in detail the possibility that space is divided
into domains that are equally likely to be made of matter or of
antimatter~\cite{CDG}. We computed the nuclear annihilation rate for
matter and antimatter near domain boundaries and showed that the
resulting relic diffuse gamma-ray flux exceeds the observed cosmic
diffuse gamma (CDG) spectrum, unless the domain size is close to that
of the visible universe.  We thus concluded that a symmetric universe with
comparable amounts of matter and antimatter is excluded, unless the
typical current size of the domains of uniform composition is $d_0>1$
Gpc\footnote{Our explicit numbers refer to a ``fiducial'' choice of
  cosmological parameters: critical mass density $\Omega =1$;
  vanishing cosmological constant $\Omega_\Lambda=0$; Hubble constant
  $H_0=75$~km/s$\cdot$Mpc or $h=0.75$.}. In this paper we study the
possibility of closing the gap, of less than one order of magnitude,
between this scale and the size of the visible universe.

Kinney {\it et al.}~\cite{CHIC} have studied ``ribbons'' in the
temperature of the cosmic background radiation (CBR) that arise at the
intersection of domain boundaries with the last scattering surface.
This assumes that matter and antimatter came into contact prior to the
time at which CBR photons last scattered. In our previous analysis we
refrained from making this very natural assumption, because it cannot
be argued to be empirically unavoidable.

In this note we {\em do} assume that matter and antimatter domains
were in contact prior to last scattering. If the effects of contact
and concomitant annihilation significantly distort the radiation from
the last scattering surface,  a single domain boundary---or
even a fraction thereof---may be detectable.  Conversely, the absence
of such signatures would complete the proof of our no-go theorem: a
universe with comparable amounts of matter and antimatter would be
excluded.

We revisit the work of Kinney {\it et al.} and reach a
different conclusion: we find that even the next generation of
satellite CBR probes will have a temperature-contrast sensitivity
inferior to what is needed to detect the effects of matter-antimatter
annihilation on the CBR.

Let $n_B(t)$ and $n_\gamma(t)$ be the baryon and photon number
densities, with $\eta\equiv n_B/n_\gamma$.  Assume spatially uniform,
equal baryon and antibaryon densities prior to last
scattering\footnote{For the conventional range of cosmological
  parameters the approximate times of last scattering and the
  formation of stable atoms are nearly coincident.}  except near
domain boundaries, where annihilation leads to depletion. To study the
effects of annihilation, we must determine the baryon annihilation
rate per unit surface $J(t)$ at the interface between a matter and
an antimatter domain. The detailed analysis of how this can be done is
found in~\cite{CDG}. Here we simply quote the results that are
relevant to the problem at hand.

Our conclusions are insensitive to the contamination of nuclear
species other than protons in the primordial plasma and to the effects
of electron-positron annihilation: we can concentrate on $p \bar p$
annihilation.  Its direct products are primarily pions ($\pi^+$,
$\pi^0$ and $\pi^-$) with similar multiplicities and energy spectra.
The end products are gamma rays from $\pi^0$ decay, energetic
electrons $e^\pm$ from the decay chain $\pi\rightarrow \mu\rightarrow
e$, and neutrinos. The behaviour of relativistic $e^+$ and $e^-$ is
sufficiently similar to justify referring to both as electrons.

The electrons from $p \bar p$ annihilation lose most of their energy
by Compton scattering off CBR photons.  The energy distribution of the
upscattered photons straddles an energy domain from the hydrogen
binding energy to a few keV, in which the K-shell photoionization
cross section is very large. Consequently, these photons keep matter
and antimatter close to a fully ionized domain boundary,
even well
after the time at which recombination would have occurred in a standard
cosmology\footnote{ Kinney {\it et al.} assume a conventional
  recombination history; at this point our analysis diverges from
  theirs.}.

Annihilation near a domain interface causes a flow to develop as new
fluid replenishes what is annihilated.  The $e^\pm$ from $p \bar p$
annihilation lose a small portion of their initial energies by
scattering off ambient electrons in the fluid. This process transfers
heat to the fluid, but prior to last scattering the effect on the matter
temperature $T(t)$ is small: CBR photons act as a large and efficient
heat bath,
with a temperature $T_\gamma(t)$ that is not significantly affected by
annihilation. Interactions between the matter and the CBR  keep the
matter temperature
$T(t)$ close to $T_\gamma(t)$.  This small increase of the electron
temperature relative to the photon temperature results in a small
distortion away from a thermal CBR spectrum, in the manner first
described by Sunyaev and Zeldovich \cite{SZ}. It is this effect,
localized along domain boundaries at last scattering, that must be
computed.

Prior to last scattering the motion of the cosmic
plasma is damped by the interaction of the ambient charged particles
with the CBR. The fluid motion is thus diffusive, described
by a time-dependent diffusion coefficient:
\begin{equation}
D_{e \gamma}(t)\equiv{45\over 4\, \pi^2\, \sigma_T\, T_\gamma^3(t)}\,,
\label{Diffcon}
\end{equation}
with $\sigma_T$ the Thompson cross section.  The diffusive nature of
the process has the welcome consequence that memory of the initial
conditions is lost as the fluid evolves. To an excellent approximation
the annihilation current $J(t)$ (also computed numerically
in~\cite{CDG}) is given by:
\begin{equation}
J(t)\equiv n_B(t) \,v(t)\simeq n_B(t) \, \sqrt{{5\,D_{e
\gamma}(t)\over 3\, \pi \, t}}\ ,
\label{diffchapuza}
\end{equation}
with $n_B(t)$ the proton number density far from the
matter--antimatter interface and $v(t)$ an effective velocity defined
here for ulterior convenience.
The current $J(t)$ is much smaller than the atomic-free-streaming
current used by the authors of~\cite{CHIC}, explaining the bulk of the
discrepancy between our conclusions and theirs.

Some of the energy carried off by electrons, $Q\sim 320$ MeV per
annihilation,
is transferred via the plasma to the CBR. Let $\Delta H_{LS}$
be the excess energy per unit volume accumulated in the CBR by the
time of last
scattering (at a certain location on the last-scattering surface) from
the effects of all previous annihilations. The conventional
Sunyaev--Zeldovich $y$-parameter describing the local distortion of
the CBR is $y=\Delta H_{LS}/(4\, u(t_{LS}))$, with $u(t_{LS})$ the CBR
energy density at the time of last scattering, $t_{LS}$.

At time $t_{LS}$ the annihilation energy is spread over a distance
orthogonal to a matter--antimatter annihilation interface of size
$\sim 2 \lambda_{LS}$, with $\lambda_{LS}$ the photon collisional
damping scale
($\lambda_{LS}\sim 15$ kpc~\cite{SILK} for our fiducial choice of
cosmological parameters).  This length scale is also comparable to the
resolution that the next generation of satellite CBR probes may
achieve.  For data acquired with this resolution, and upon neglect of
geometrical factors such as the inclination at which the last
scattering surface intersects a given domain interface, we may
estimate the $y$-parameter distorsion of the ``ribbon''. The result:
\begin{equation}
y\sim{1\over 8}\,{Q\, n_B(t_{LS})\over \lambda_{LS}\,u(t_{LS})}\;
\int^{t_{LS}} \bar f(t)\,v(t)\,dt\,
\label{y}
\label{SZ}
\end{equation}
is dominated by times close to $t_{LS}$. Here $\bar f(t)$ is the
average fractional energy deposition in the CBR by annihilation
electrons~\cite{CHIC}. To complete our calculation we must compute $f(t)$.

An average of nearly four electrons is made per $p \bar p$
annihilation, with an energy spectrum peaked at $E_e\equiv m_e\gamma_e
\sim 80$ MeV. Prior to recombination, Compton scattering on CBR
photons completely dominates over the other electron energy-loss
mechanisms (red-shifting and scattering on ambient matter). The
spectrum of photons Compton-up-scattered by a single annihilation
electron~\cite{CDG} at time $t$ may be approximated by:
\begin{equation}
{dn\over dw} \simeq  {8\, E_e \over 3\, \pi\,\gamma_e\;
\sqrt{3\,T_\gamma\,w^3}}\;
\left[1-{w\over 3\,\gamma_e^2\,T_\gamma}\right]^{3/2}
\Theta[3\,\gamma_e^2\,T_\gamma-w]\, .
\label{spectrum}
\end{equation}
This approximation is extremely good for the photon energies
$w>10T_\gamma(t)$ that dominate our results\footnote {In~\cite{CHIC}
  a spectrum with a fixed energy $3\gamma_e^2\,T_\gamma$ is used.
  This results in an overestimate of the energy-deposition efficiency
  by a factor $\sim 5$.}.

Red-shift and Compton scattering on ambient electrons have comparable
effects in making a Comptonized photon lose energy:
\begin{equation}
{dw\over dt}+H(t)\, w=-\,{w^2\over m_e}\;\sigma_T\,n_e(t)\ ,
\label{lose}
\end{equation}
with $H(t)$ the Hubble expansion rate and $n_e(t)$ the electron number
density $n_e\simeq n_0\,z(t)^3$ (for the large red-shifts of interest
we do not make the distinction between $z$ and $1+z$).  Let
$z_{LS}\sim 1100$ be the last-scattering red-shift.  At that time, a
photon made at an earlier epoch $z$ would, if it did not interact,
have an energy $w\,z_{LS}/z$, but because of the interaction described
by the rhs of Eq.~(\ref{SZ}) it has a lower energy. The difference
 between these two energies, $w_H (w,z)$, is the photon's contribution to
the excess energy density $\Delta H_{LS}$ relevant to the CBR spectral
distortion.  Solving Eq.~(\ref{lose}) we find:
\begin{eqnarray}
 w_H (w,z)  & = & w\,{z_{LS}\over z}\; f(w,z)\nonumber\\
  f(w,z) & \equiv &  {a \, w \left(z^{5/2}-z_{LS}^{5/2}\right)
\over z+ a \, w \left(z^{5/2}-z_{LS}^{5/2}\right)}
\nonumber\\
  a  & \equiv & {2\,\sigma_T\, n_0 \over 5\, m_e\, H_0}\;.
\label{effic}
\end{eqnarray}
The mean efficiency $\bar f$ in Eq.~(\ref{SZ}) is the energy-weighted
average of $f(w,z)$
over the photon spectrum of Eq.~(\ref{spectrum}):
\begin{equation}
\bar f (z) = {1 \over E_0 } \; \int_{0} ^{\infty}
\,f(w,z)\, w\,{dn \over dw} \, dw\ .
\label{ave}
\end{equation}

Performing the integrations in Eqs.~(\ref{SZ}) and (\ref{ave}) we obtain
$y\sim 3.4 \times 10^{-7}$.  The result scales roughly as $\eta^2/H_0$ and
we have used $\eta=6 \times 10^{-10}$, $H_0=50$~km/s/Mpc to illustrate the
maximal expectation, within errors.

We have neglected a small additional energy deposition by Comptonized
photons.  Regions lying far from domain boundaries recombine as in a
standard cosmology. A moving front develops between ionized and
recombined regions as the photon flux progresses, depositing a
fraction of its energy as it reionizes the medium.  The velocity of
the front is $v_f \sim {c/ (1+\xi})$ where $\xi$ is the ratio of the
atomic number density to that of the incident photon flux. Around
recombination the details of this process are complicated, and rather
than attempting a detailed description, we notice that the energy is
deposited over distances larger than the collisional damping scale
$\lambda_{LS}$. Consequently, an absolute upper bound to the
contribution to a $y$-distorsion can be obtained by using
Eq.~(\ref{SZ}) (in which the energy is distributed in a region of
width $\lambda_{LS}$) with an assumed energy-deposition efficiency
$\bar f(t)=1$. The bound scales as $\eta^{3/2}/H_0$ and has a value
$y=1.4 \times 10^{-6}$ for $\eta=6 \times 10^{-10}$, $H_0=50$ km/s/Mpc.

Our result for the temperature non-uniformity along a matter--antimatter
interface is well below the sensitivity levels of currently planned
observations.
In constraining a baryon-symmetric cosmology,
the exquisite detail with which the CBR can be
studied is no match for a rough measurement
of the diffuse gamma ray background~\cite{CDG}.

\end{document}